\documentclass[a4paper]{article}
\usepackage{cite}
\usepackage{amsmath,amssymb,amsfonts}
\usepackage{algorithmic}
\usepackage{graphicx}
\usepackage{balance}
\usepackage{textcomp}
\usepackage{xcolor}
\usepackage{lscape}

\begin{document}

\title{A Machine Learning-based Recommendation System for Swaptions Strategies}

\author{Adriano Soares Koshiyama, Nick Firoozye and Philip Treleaven\\
	\textit{Department of Computer Science} \\
		\textit{University College London}\\
		London, United Kingdom \\
		adriano.koshiyama.15@ucl.ac.uk
}

\maketitle

\begin{abstract}
Derivative traders are usually required to scan through hundreds, even thousands of possible trades on a daily basis. Up to now, not a single solution is available to aid in their job. Hence, this work aims to develop a trading recommendation system, and apply this system to the so-called Mid-Curve Calendar Spread (MCCS), an exotic swaption-based derivatives package. In summary, our trading recommendation system follows this pipeline: (i) on a certain trade date, we compute metrics and sensitivities related to an MCCS; (ii) these metrics are feed in a model that can predict its expected return for a given holding period; and after repeating (i) and (ii) for all trades we (iii) rank the trades using some dominance criteria. To suggest that such approach is feasible, we used a list of 35 different types of MCCS; a total of 11 predictive models; and 4 benchmark models. Our results suggest that in general linear regression with lasso regularisation compared favourably to other approaches from a predictive and interpretability perspective.
\end{abstract}

\section{Introduction}

Derivative traders are usually required to scan through hundreds, even thousands of possible trades on a daily basis. A concrete case is the so-called Mid-Curve Calendar Spread (MCCS), a derivatives package that involves selling an option on a forward-starting swap and buying an option on a spot-starting swap with longer expiration \cite{Book:CorbIRS:2012,Book:NatenbergOption:2014}. In such a package, traders look for the historical carry and the breakeven width levels, metrics that can be easily inferred from the terminal or aged payoff profile of the MCCS, shown in several heatmaps made by the research team. After that, they rank the most prominent ones to offer a client or to proceed in some proprietary trading. In general, the straightforwardness and swiftness that the decisions are made is the main upside of this framework.

However, one might notice that the main downsides of such approach are: (i) substantial information on the underlying like sensitivities, implied volatility, etc. are usually not taken into account; (ii) using the previous example, high historical values for carry and breakeven widths are more necessary rather than sufficient conditions for a profitable MCCS trade, being such argument extensible to other trades as well; (iii) a trader can quickly judge if an individual trade is worthwhile to invest, but may take some time to find it; and (iv) after a given period, traders tends to only look at a small subset of possible trades (small area on the heatmap), rather than the all available selection. Hence, a systematic approach where more information at hand is crossed and aggregated to find good trading picks and undoubtedly increase the trader's productivity.

Therefore, the objective of this work is to develop a trading recommendation system that can aid derivatives traders in their day-to-day routine. Being more specific, our solution is based on the following pipeline: (i) on a certain trade date, we compute metrics and sensitivities related to MCCS; (ii) these metrics are feed in a model that can predict its expected return for a given holding period; and after repeating (i) and (ii) for all trades we (iii) rank the trades using some dominance criteria. Our final solution is a model-based heatmap with the attractiveness scores for each trade, which can be offered to the traders and salespeople on a daily basis.

In this sense, we organised this work as follows: next section presents a literature review on existing approaches to return/price prediction/estimation in different areas and instruments, as well as a brief description on MCCS trades. The third section displays the dataset that comports the MCCS trades, showing how the information is computed and gathered, which variables are the input and outputs, and the main assumptions that are embedded in it. Then, we move to modelling strategy, highlighting the main models that are going to be used as candidates for the recommendation system, how they are tested and have their performance assessed. Finally, we exhibit the results and discussions, closing this work with some concluding remarks and future directions for research.

\section{Background}

\subsection{Related Works}

Literature provides a growing body of evidence that price changes can be predicted, that is, in particular circumstances and periods securities violate the Efficient Market Hypothesis \cite{Book:CampbellFinEconometrics:1997,Paper:MalkielEfficient:2003}. In this sense, researchers have employed different modelling approaches and information sets to predict price changes across a range of assets. When we scan the literature for cash instruments (equities, bonds, foreign exchange, etc.) focused only in using past returns as the main source for prediction, we can find works that tap into Bayesian forecasting \cite{Paper:ZhouBayesReturnForecasting:2014}, Nonparametric Predictive Inference \cite{Paper:BakerReturnForecastNPI:2016}, Forecasting Combination \cite{Paper:ElliottReturnForecastingCombination:2013}, Generalized Exponential Weighted Moving Average \cite{Paper:NakanoReturnForecastingEWMA:2017}, Support Vector Machines (SVM) \cite{Paper:KarathanasopoulosForecastingReturnsSVM:2016}, Shallow and Deep Neural Networks architectures \cite{Paper:GerleinForecastingReturnsNNClassif:2016,Paper:ChongReturnForecastingDeepLearning:2017,Paper:ZhouReturnForecastingDendriticNeuron:2016,Paper:DengReturnForecastingDeepRL:2017}, Random Forest and Gradient Boosting Trees \cite{Paper:KraussReturnForecastingRandomForest:2017}, and so forth. The list of proposed methodologies keeps growing, in which equities or indices appears as the dominant asset class to apply these algorithms. Collectively, they provide evidence that some forecastability over returns can be achieved by putting in place complex models with a suitable training scheme.

Contrasting with the emphasis that researchers in cash instruments put on return predictability, when we devote our attention to research in derivatives instruments (options, swaps, swaptions, etc.) it is clear that most of the effort is concentrated on pricing these contracts. In parallel to the traditional framework, alternative ways of pricing and trading started to emanate relying on fewer assumptions and more data-driven. We can pinpoint approaches that use Neural Networks for option pricing and hedging with daily S\&P 500 index daily call options \cite{Paper:GencayPricingNN:2001} as well as for real-time pricing and hedging options on currency futures of EUR/USD at tick level \cite{Paper:SpreckelsenPricingHedgingNN:2014}. It is worth to mention other approaches in the derivatives realm, such that the prediction of pricing and hedging errors for equity-linked warrants with Gaussian Process models \cite{Paper:HanPricingHedgingGaussianProcess:2008}, building machine learning models for predicting option prices over KOSPI 200 Index options \cite{Paper:ParkPricingMachineLearning:2014} and a general study on forecasting option price distributions using Bayesian kernel methods \cite{Paper:ParkForecastingOptionPrice:2012}.

When we devote our attention to the asset type that this work is dedicated, interest rate swaptions, a similar pattern persists: most of the research is related to pricing and not to return prediction. Regarding pricing, the same tradition of relying on stochastic calculus techniques is followed \cite{Book:BrigoInterestRateModels:2007, Book:RebonatoSABR:2011}. Regarding potential alternatives using more data-driven approaches as we saw with currency, indices and equities options, we can only mention the work of Souza et al. \cite{Paper:BelezaSouzaMLVasicekModel:2012} which calibrates the Vasicek interest rate model under the risk neutral measure by learning the model parameters using Gaussian processes for regression. Considering trading strategies and return prediction, we can find even less academic research, being perhaps most of the research residing inside the counterparts that exchange such products (banks, hedge funds, etc.). This shortage of published research might be linked with the absence of ready to use and publicly available datasets, similar to the ones found in cash products since these instruments are traded off-exchange.

Based on this review of existing approaches to return/price prediction/estimation in different areas and instruments, to the best of our knowledge, our work is the first attempt to build a trading recommendation system in the context of derivatives. Our approach is not the only novel from a modelling perspective, but instead of trading the vanilla product (receiver/payer interest rate swaption), we prefer to focus on options strategies (calendar spreads, straddles, etc.) which in many cases is the package that is in practice traded. By thinking in terms of the package, in this case, a Mid-Curve Calendar Spread, rather than the individual constituents we unlock some features that can only be computed in this situation, like the carry at expiry, breakeven width and so on.

Therefore, we can train our models not only using past returns but also using sensitivities as well as information derived from the package payoff function. By portraying in this manner our investment strategy, we have a large information set that can substantially add information to aid forecasting returns. But as a counter-effect, this poses a new challenge on separating relevant features in a dynamic context. In this respect, the combination of temporal cross-validation, a diverse set of models and regularisation/feature selection can provide a robust framework for trading strategies backtesting and assessment. But before presenting such framework, next section gives a brief view on MCCSs trades.

\subsection{Mid-Curve Calendar Spreads}

\begin{figure}[h!]
	\centering
	\includegraphics[width=\linewidth]{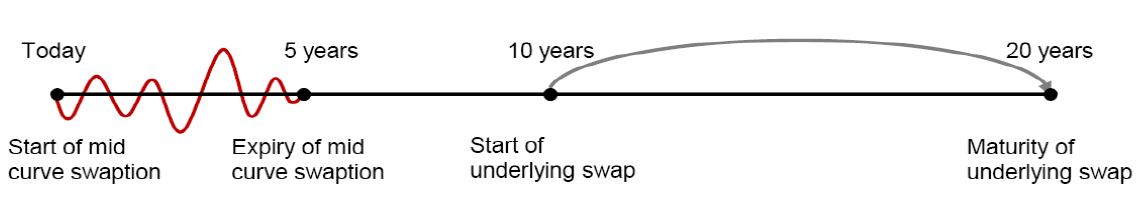}
	\caption{Mid-curve Swaption: 5y mid-curve on 5y10y swap rate -- the volatility of a forward-starting swaption, called mid-curve, whose strike is set at inception and but the underlying swap starts several years following the option expiry date. }
	\label{MidCurveSwaptionExample}
\end{figure}

Mid-Curve Calendar Spread (MCCS) is a package involving short selling an option on a forward-starting swap and going long a longer-expiry swaption on the same underlying swap \cite{Book:CorbIRS:2012}. There is a counterpart with many similarities for equities -- check in \cite{Book:NatenbergOption:2014} for more information. Investors typically use MCCS to take a view on forwarding volatility. This comes from the fact that, conceptually, spot volatility can be decomposed into forward volatility and mid-curve volatility. Taking 10y10y\footnote{This notation is extensively used during this work. In this case, the first 10y means a spot swaption with 10 year of expiration, while the second 10y refers to the swap tenure.} for example, Figure \ref{MidCurveSwaptionExample} illustrate the time periods covered by different interest rate volatilities and their instruments. The red lines indicate the time over which interest rate volatility exposure is taken, and the grey line indicates the underlying forward swap rate.

\begin{figure}[h!]
	\centering
	\includegraphics[width=0.75\linewidth]{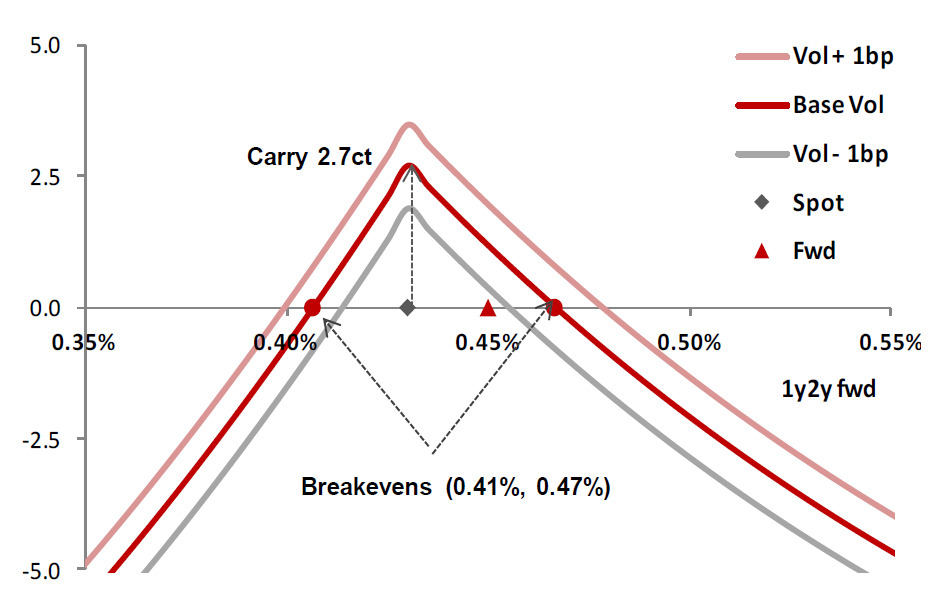}
	\caption{Payoff profile for an EUR 1m1y2y. } \label{MCCSPayoff1m1y2y}
\end{figure}

Figure \ref{MCCSPayoff1m1y2y} presents the payoff profile for an EUR 1m1y2y\footnote{Short selling a 1m1y2y mid-curve swaption and going long a longer-expiry 13m2y spot swaption.}. 

We plot the payoff profiles for current volatility and up and down volatility scenarios, noting that the long vega position means that the payoff profile shifts up in a rising volatility environment and correspondingly shifts down in a falling vol environment. We calculate the (volatility adjusted) breakevens as being $0.41\%-0.47\%$, giving little protection against selloffs. We note that forwards in a $\pm 1$ volatility band leave them at $0.40\%-0.48\%$, a range just marginally larger than our breakeven range (i.e., the trade should pay off just slightly less than $66\%$ of the time). MCCS can result in what we think of as turbocharged carry, primarily because of the risks that they have (which fortunately can be balanced with the returns in a way which results in relatively attractive trades). Based on these characteristics of MCCS, next section presents how we elaborated the methodology to build this trading recommendation system. 

\section{Methodology}

In summary, our solution develops the following roadmap (also schematically described in Figure \ref{TradingRecSys}): 

\begin{figure*}[h!]
	\centering
	\includegraphics[width=\linewidth]{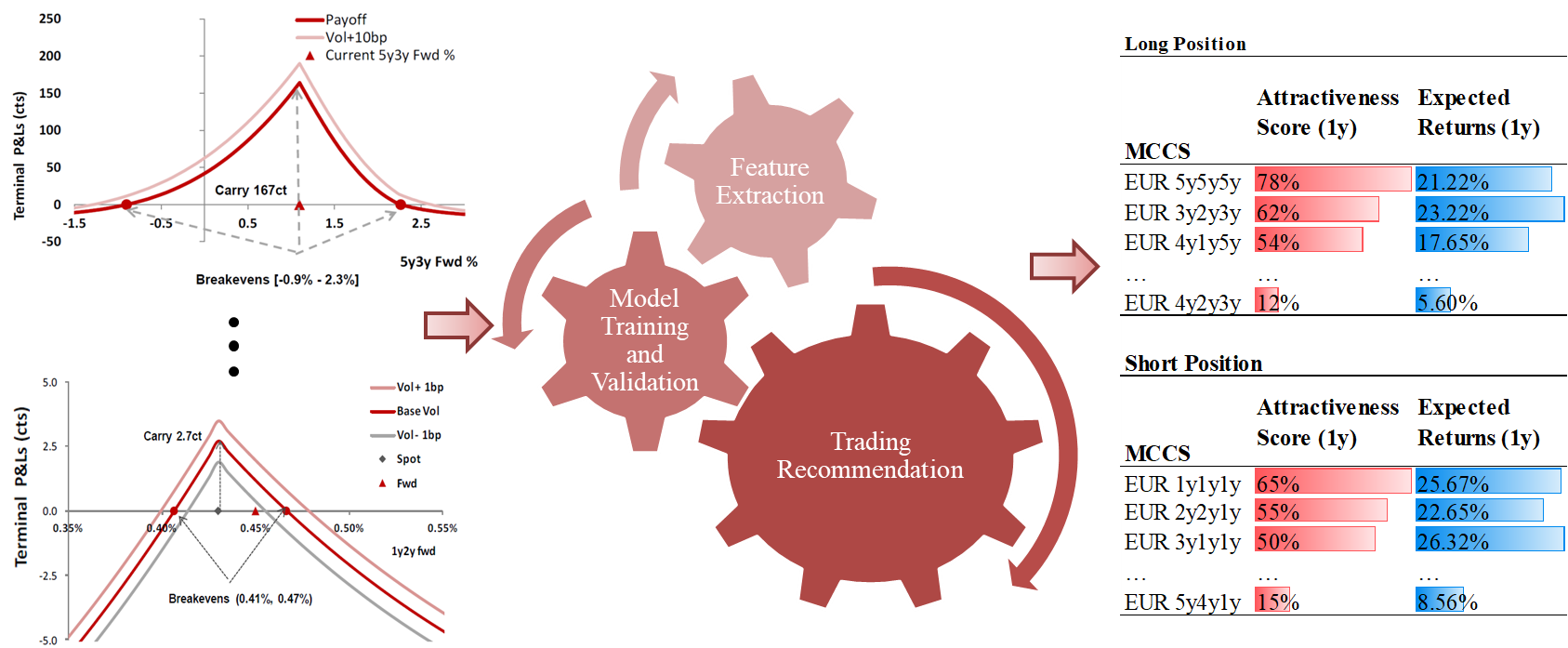}
	\caption{Flowchart describing the input-output schemes from the proposed trading recommendation system for MCCS trades.} \label{TradingRecSys}
\end{figure*}

\begin{enumerate}
	\item \textbf{Data}: On a certain trade date, we \textbf{calculate metrics and sensitivities} related to an MCCS package; 
	\item \textbf{Modelling}: These metrics are \textbf{feed in a predictive model} that outputs its expected return for a given holding period (e.g., one year); 
	\item \textbf{Recommendation}: After repeating (i) and (ii) for all MCCS we (iii) rank them based on the expected returns using some criteria. 
\end{enumerate}

Following this outlined structure, the next three subsections describe in more details when and which MCCS trades were recorded (Dataset), which predictive models were trained and how they were assessed (Modelling) and how the long/short trading signal is computed for each MCCS (Recommendation). Finally, last subsection presents which metrics were used to evaluate the recommendation system performance when a certain predictive model candidate is underpinning it.

\subsection{Dataset}

During our experiments, we opted to use the trades displayed in Table \ref{MCCSTrades}. Although many other configurations are available in practice, these are the ones with longest historical data available, which is important when it is necessary to fit a predictive model. As it can be seen, all trades are in Euro, ranging from different expiries (1y-5y), forwards (1y-5y) and swap tenures (1y-5y and 8y).

\begin{table*}[h!]
	\centering
	\caption{Configuration of the MCCS trades used.} \label{MCCSTrades}
	\scriptsize
	\begin{tabular}{cccc|cccc}
		\hline
		\hline
		Currency & Expiry & Forward & Swap & Currency & Expiry & Forward & Swap \\
		\hline
		EUR & 1y & 1y & 1y & EUR & 3y & 3y & 2y \\
		EUR & 1y & 1y & 4y & EUR & 3y & 4y & 1y \\
		EUR & 1y & 2y & 3y & EUR & 3y & 5y & 5y \\
		EUR & 1y & 2y & 8y & EUR & 4y & 1y & 1y \\
		EUR & 1y & 3y & 2y & EUR & 4y & 1y & 4y \\
		EUR & 1y & 4y & 1y & EUR & 4y & 2y & 3y \\
		EUR & 1y & 5y & 5y & EUR & 4y & 2y & 8y \\
		EUR & 2y & 1y & 1y & EUR & 4y & 3y & 2y \\
		EUR & 2y & 1y & 4y & EUR & 4y & 4y & 1y \\
		EUR & 2y & 2y & 3y & EUR & 4y & 5y & 5y \\
		EUR & 2y & 2y & 8y & EUR & 5y & 1y & 1y \\
		EUR & 2y & 3y & 2y & EUR & 5y & 1y & 4y \\
		EUR & 2y & 4y & 1y & EUR & 5y & 2y & 3y \\
		EUR & 2y & 5y & 5y & EUR & 5y & 2y & 8y \\
		EUR & 3y & 1y & 1y & EUR & 5y & 3y & 2y \\
		EUR & 3y & 1y & 4y & EUR & 5y & 4y & 1y \\
		EUR & 3y & 2y & 3y & EUR & 5y & 5y & 5y \\
		EUR & 3y & 2y & 8y &  &  &  &  \\
		\hline
		\hline
	\end{tabular}
\end{table*}

For each configuration, at time $t$ we agree with a counterpart to trade this package using the At the Money Forward (ATMF) rate as the strike, paying or receiving the present value $PV_t$. The $PV_t$ is computed via SABR model \cite{Book:RebonatoSABR:2011}, using information and parameters (e.g., spot, forward rates and rate-rate correlation) calibrated using market data on a daily basis. From the same model that computed the $PV_t$, we can also obtain other metrics and sensitivities as those displayed in Table \ref{MCCSFeatures}.

\begin{table*}[h!]
	\centering
	\caption{Metrics and sensitivities computed for each available package at time $t$.} \label{MCCSFeatures}
	\footnotesize
	\begin{tabular}{cc}
		\hline
		\hline
		\multicolumn{2}{c}{Features} \\
		\hline
		$PV$ & Strike \\
		Carry at Expiry (Carry) & Breakeven Width (BE Width) \\
		Aged 1y Carry & Theta \\
		ATMF Implied Volatility (Implied Vol) & Gamma \\
		Vega & Curve Carry (Aged 1y) \\
		Time Carry (Aged 1y) &	Volatility Carry (Aged 1y) (Vol Carry) \\
		\hline
		\hline
	\end{tabular}
\end{table*}

Carry and BE Width are those obtained looking at the payoff profile at expiry. The Aged 1y Carry is produced by ageing the trade by one year (moving closer to the expiration) and estimate the payoff profile computing the carry. Theta, Vega and Gamma are the sensitivities of the instruments by a change in time, volatility and a wider range of underlying rate movements, respectively. These and the ATMF Implied Vol are backed by the SABR model too. Curve, Time and Volatility Carry are the amount of Aged 1y Carry that can be attributed to the changes in certain sensitivities from spot to forward, such as the Delta (Curve), Theta (Time) and Vega (Volatility). These can also be used as tools to understand which factors most influence the instrument value over time. 


After computing all these metrics at time $t$, we hold the trade until $t+h$ where $h$ can be two weeks, one month, one year, and so on, as long as $t+h$ is before or at expiration. In time $t+h$ we compute the $PV_{t+h}$ of the same trade again, using the new economic scenario available (e.g. rates, change in model parameters). By agreeing on buying back or selling the current trade for $PV_{t+h}$ we can compute the Holding k-period Return of the trade started at time $t$ by:
\begin{equation}
R_{t}^{(h)} = \frac{PV_{t+h} - PV_t}{PV_t}
\end{equation}

In summary, Table \ref{MCCSDataset} presents an example of information in a wide format that is available when we combine the data from time $t$ and $t+h$.

\begin{table}[h!]
	\centering
	\caption{Example of information available at time $t$ and $t+h$ for the MCCS.} \label{MCCSDataset}
	\footnotesize
	\begin{tabular}{c|cccc|c}
		\hline
		\hline
		Instant ($t$) & $PV_t$ & $R_{t}^{(h)}$ & $R_{t-h-1}^{(h)}$ & Strike & Features \\
		\hline
		1 & 300 & 0.3 & - & 3.4 & ... \\
		2 & 320 & 0.2 & - & 3.2 & ... \\
		... & ... & ... & ... & ... & ... \\
		$t+h-1$ & 250 & -0.1 & - & 1.8 & ... \\
		$t+h$ & 260 & -0.05 & - & 1.9 & ... \\
		$t+h+1$ & 270 & -0.2 & 0.2 & 2.0 & ... \\
		... & ... & ... & ... & ... & ... \\
		$T$ & 250 & - & 0.1 & 2.2 & ... \\
		\hline
		\hline
	\end{tabular}
\end{table}

Note that in the most contemporaneous period (close to $T$) we do not have the $PV_{T+h}$ and so, we cannot compute $R_{T}^{(h)}$. Conversely, if we want to use lagged returns $R_{t-h-1}^{(k)}$ as explanatory forces for $R_{t}^{(h)}$, then in the beginning this information is also not available. Therefore, our dataset is trimmed at the beginning and the end mainly by the value of $h$. If $h$ is small, such as two weeks or one month, the trimming is imperceptible and, therefore, may not affect the model fitting and validation. However, if $h$ is large such as two or three years, this might reduce the samples available substantially, decreasing the range of models and cross-validation schemes that might be employed for this task. Based on these procedures, metrics and observations, Table \ref{MCCSParametersDataset} express other details that we used during our experiments to generate the dataset.

\begin{table}[h!]
	\centering
	\caption{Details used to generate the MCCS trade dataset.} \label{MCCSParametersDataset}
	\footnotesize
	\begin{tabular}{c|c}
		\hline
		\hline
		Detail & Value \\
		\hline
		Period & September 2006 to September 2016 \\
		Holding Period ($h$) & 1 year \\
		Trade Frequency & Weekly (usually on Wednesday) \\
		Strike & At the Money Forward (ATMF) \\
		Lagged data ($h-p$) & $p = 1, 2$ and $3$ lagged returns \\
		\hline
		Assumption & Characteristic \\
		\hline
		Bid-Ask & Middle Rate \\
		Transaction Costs & Entry and Unwind = $0.75 \times Vega_t$ \\
		Funding Rate & Libor 3 month rate \\
		\hline
		\hline
	\end{tabular}
\end{table}

Therefore, we gathered data from trades entered on a weekly basis from September 2006 to September 2016. These trades are struck ATMF, using the $PV_t$ computed from the Middle Rate (in practice, some bid-ask spread would be imbued proportional to the Vega). After holding for one year ($h=1y$) the trade, we compute the arithmetical returns that are, therefore by definition, automatically annualised. These returns are gross, and so we need to take into account the transaction costs (hedging costs and fixed fees charged by the derivatives desk) as well as some future funding rate. These values are also outlined in Table \ref{MCCSParametersDataset}, where the transaction costs of 0.75 as a fraction of Vega were chosen not only to taken into account the transaction cost, but also some potential bid-ask spread on the start/unwind of the trade. The 3-month London Interbank Overnight Rate (LIBOR) was chosen as the funding cost/benchmark rate to compute excess returns.

Using these assumptions, next subsection presents the modelling strategy that taps into this dataset to create the recommendation system for the MCCS trades.

\subsection{Modelling}

In relation to modelling, our general model is a system of uncoupled equations:
\begin{eqnarray}
R_{t, 1}^{(1y)} = f_1(features_{t, 1}) + \varepsilon_{t, 1} = \hat{R}_{t, 1}^{(1y)} + \varepsilon_{t, 1}\\
R_{t, 2}^{(1y)} = f_2(features_{t, 2}) + \varepsilon_{t, 2} = \hat{R}_{t, 2}^{(1y)} + \varepsilon_{t, 2} \\
... \nonumber \\
R_{t, n}^{(1y)} = f_n(features_{t, n}) + \varepsilon_{t, n} = \hat{R}_{t, n}^{(1y)} + \varepsilon_{t, n} 
\end{eqnarray}

where for each MCCS trade ($i=1,...,n$) there is an i-th predictive model $f_i$ that is feed with a set of pre-calculated features (BE Width, Carry, etc.) and returns an estimate of the holding 1y-period return $\hat{R}_{t, i}^{(1y)}$. As the model is an approximation, some noise/error is expected, and in the modelling aspect, this is expressed as the $\varepsilon_{t, i}$ component. After defining which variable is intended to be predicted, the remaining points are: which models are available to embody $f_i$ and how the fitting, validation and selection of these models are going to be made.

About the first point, in the first rows of Table \ref{MCCSParametersModel} we display the models that we used during our experiments, with their mathematical descriptions and usage found in the following references \cite{Book:BishopML:2007,Book:Duda:2000,Book:HastieElements:2016,Book:HaykinNeural:2009}.

\begin{landscape}
	
	\begin{table*}[h!]
		\centering
		\caption{Parameters used to model the MCCS trade dataset.} \label{MCCSParametersModel}
		\scriptsize
		\begin{tabular}{c|c|c|c}
			\hline
			\hline
			\textbf{Abbreviation} & \textbf{Model} & \textbf{Fixed Hyperparameters} & \textbf{Cross-Validated Hyperparameters} \\
			\hline
			Classic Regression & Classical Linear Regression & None & None \\	
			BackSel Regression & Stepwise Regression & Backward Selection & None \\
			Ridge Regression & Ridge Regression & None & $\lambda = \{10^0, 10^{-1}, 10^{-1}, 10^{-2}, 10^{-3}\}$ \\
			Lasso Regression & Lasso Regression & None & $\lambda = \{10^0, 10^{-1}, 10^{-1}, 10^{-2}, 10^{-3}\}$ \\
			KRR-RBF & Kernel Ridge Regression & Radial-Basis Function kernel & $\lambda = \{10^0, 10^{-1}, 10^{-1}, 10^{-2}, 10^{-3}\}$ \\
			& & & and $\gamma = \{10^{-2}, 10^{-1}, 10^{0}, 10^{1}, 10^{2}\}$ \\
			kNN & k-Nearest Neighbours & Euclidean Distance &  $k = \{3, 5, 7, 9\}$ \\
			CART & Classification and Regression Tree & MSE Function & Max depth $= \{2, 3, 5, 7\}$  \\
			Random Forest & Random Forest & Number of trees $= 333$ and Max depth $= 5$ & None \\
			Grad Boost Reg & Gradient Boosting Tree & Number of trees $= 333$ and Max depth $= 5$ & Learning Rate: $\{0.1, 0.3, 0.5\}$ \\
			MLP & Multi-Layer Perceptron & Single hidden layer with hyperbolic & $\lambda = \{10^0, 10^{-1}, 10^{-1}, 10^{-2}, 10^{-3}\}$ \\
			& & tangent as transfer function & and $nn = \{5, 7, 12\}$ \\
			SVR-RBF & Support Vector Regression & Radial-Basis Function kernel & $C = \{10^0, 10^{1}, 10^{2}, 10^{3}\}$ \\
			& & & and $\gamma = \{10^{-2}, 10^{-1}, 10^{0}, 10^{1}, 10^{2}\}$ \\
			
			\hline
			\textbf{Abbreviation} & \textbf{Baseline Model} & \multicolumn{2}{c}{\textbf{Parameter}} \\
			\hline
			Mean Pred & Average Prediction & & \\
			Naive & Naive Model & & \\
			Z-Score: BE-Width & BE Width feature & Rolling window of size = 1 year & $S_t = \lfloor_{-1}((Z > 1)*Z)/(3)\rceil^{+1}$ \\
			Z-Score: CarryAtExpiry & Carry at Expiry feature & Rolling window of size = 1 year &  $S_t = \lfloor_{-1}((Z > 1)*Z)/(3)\rceil^{+1}$ \\
			\hline		
			& \textbf{Other Parameters} & \multicolumn{2}{c}{\textbf{Values}} \\
			\hline
			& Warm-up Period ($L$) & $L_{outer} = 2$ years & $L_{inner} = 1$ year \\
			& k-rolling-cv & $k_{outer}=1$ week for outer & $k_{inner} \approx (T_{train}-L_{inner})/5$ for inner \\
			& Outlier Treatment & Winsorizing & \\
			& Winsorizing Quantiles & 0.01 and 0.95 & \\
			& Missing Data Treatment & Remove & \\
			\hline
			\hline
		\end{tabular}
	\end{table*}

\end{landscape}

In Table \ref{MCCSParametersModel} Model column presents a plethora of models that this work has fitted for this prediction purpose: we started from simple predictive models such as Classical Linear Regression, k-Nearest Neighbours and Classification and Regression Tree, towards those that can seamlessly exhibit nonlinear behaviours, like Random Forest, Kernel Ridge Regression, Multi-Layer Perceptron and Support Vector Regression. Some of these methods had their hyperparameters held constant across all experiments (Fixed Hyperparameters column), or because we wanted to apply a particular form of a method (RBF kernel, single hidden layer, etc.) or because during a warm-up phase we noticed that they did not affect substantially the results (hyperbolic tangent, increasing number of trees, etc.).

For certain models, the Cross-Validated Parameters column shows which hyperparameters were optimised before the prediction step. For instance, suppose the case of Ridge Regression and the need to define the regularisation value ($\lambda$) appropriately. Consider that we have a set of training pairs $(features_t, R_{t}^{(1y)})_{t=1}^{L}$ of size $L$, and for this sample we subset it in k-rolling-cross-validation (k-rolling-cv) folders (better explained later in this subsection). Then, we train and test using this scheme the Ridge Regression model with one of the predefined $\lambda$, say $\lambda = 10^0$. We compute some performance function on the test set (Mean Squared Error -- MSE) and repeat this process for all $\lambda$ values available. We use in the final model the $\lambda$ that on average had the lowest MSE.

We fitted usual benchmarks found in the literature for regression and forecasting modelling: the Average and Naive models \cite{Book:HastieElements:2016,Book:HyndmanETS:2008}. We also implemented the benchmarks that traders use to assess whether a particular MCCS is worth to be pitched or traded: BE Width and Carry at Expiry. We replicated the way traders look to these features, by computing z-scores\footnote{a z-score is defined by: $Z-score = \frac{X - \mu}{\sigma}$ where $X$ represent the actual value of a certain variable, $\mu$ and $\sigma$ the average and standard deviation of $X$ in a period.} based on average and standard deviation on rolling window of size equal to 1 year. The signal for going long/short is done by a thumb rule with a simple rationale: if a certain metric has a z-score above or equal to $\pm 3$, the trader goes fully long (+)/short(-) in the trade, since it is a very extreme event. Otherwise, it reduces the leverage on it, until it below one standard deviation of distance from the rolling average.

We removed any missing data, and clipped extremes values, mainly in returns above the 95\% percentiles (in our case it can be due to some numerical problems, or some extreme scenarios related to 2008-2009 financial crisis period). Next subsection presents the final component of our roadmap: recommendation system.

\subsection{Recommendation}

The recommendation of a certain trade can be made solely on some normalised version of the expected return for holding 1y-period the i-th trade ($\hat{R}_{t, i}^{(1y)}$). Given that each model will be providing individual forecasts for each MCCS and after that their performance will be assessed locally and globally, a more suitable manner to proceed would be to assign a credit based on the tracking record of a model to predict a particular MCCS trade. Hence, we will be weighted up or down a signal not only based on the magnitude of a model prediction but also by its quality. Then, consider as $\hat{R}_{t, i}^{(1y)}$ the expected return for holding 1y-period the i-th trade. Now, define the new signal function $S_{t, i}$ by:

{\footnotesize
	\begin{eqnarray}
	S_{t, i} = \frac{\hat{R}_{t, i}^{(1y)} \times Rho_{\hat{R}_{t, i}^{(1y)}, R_{t, i}^{(1y)}}}{\max(|\hat{R}_{t, i}^{(1y)} \times Rho_{\hat{R}_{t, i}^{(1y)}, R_{t, i}^{(1y)}}|, ..., |\hat{R}_{t-h, i}^{(1y)} \times Rho_{\hat{R}_{t-h, i}^{(1y)}, R_{t-h, i}^{(1y)}}|, 0)} \nonumber
	\end{eqnarray}
}

where the strength of the i-th long/short signal is given by its expected return, scaled by the maximum weighted return that a long/short position on the same trade (that is why the returns are in absolute terms) was expected to yield in the previous h-period (in this case 1 year). Therefore, the trade with the maximum weighted return in absolute terms will have $|S_{t, i}| = 1$ as well as those close to zero will yield $S_{t, i} \approx 0$. The weight/credit of a certain prediction is based on the historical Pearson correlation coefficient, that is, adherence between the actual and predicted values.

\subsection{Evaluation Metrics}

Below we outline two types of metrics: one that focuses on the predictive performance that the model provided, and other three that are based on the profit/loss that its application harvested during the backtest. Set by $R_t^{(S)} = R_t^{(1y)} \times S_t(\hat{R}_t^{(1y)})$ the strategy return (combination of the realized/observed excess returns and the signal -- function of a model prediction)\footnote{For the sake of brevity we dropped the subscript that refers to a particular trade ($i$).}, we can compute the following metrics:

\begin{itemize}
	
	\item Pearson Correlation Coefficient (Rho): it is a dimensionless measure of the linear dependence between the actual and predicted values:
	\begin{equation}
	Rho = \frac{Cov[R_t^{(1y)}, \hat{R}_t^{(1y)}]}{\sqrt{Var[R_t^{(1y)}]Var[\hat{R}_t^{(1y)}]}}
	\end{equation}
	where $Cov$ and $Var$ are the covariance and variance operators. It ranges from $[-1, +1]$, with $-1$ representing a perfect inverse linear association, and $+1$ the opposite. In our case, we benefit more when $Rho$ is close to $+1$. In the context of linear models, a higher predictive power is a necessary condition for profitable trades (see \cite{Book:AcarTradRules:2002}), hence by minimising the predictive error we are somewhat trailing a path for profits maximisation, albeit such causation is not very clear since this is not a sufficient condition.
	
	\item Average Return (Avg Return): is the arithmetic average of the strategy returns:
	\begin{equation}
	\bar{R}^{(S)} = \frac{\sum_{t=1}^{T} R_t^{(S)}}{T}
	\end{equation}
	
	\item Standard Deviation: is the estimator of the dispersion around the strategy average returns (a risk measure in certain sense):
	\begin{equation}
	\sigma_{R^{(S)}} = \sqrt{\frac{\sum_{t=1}^{T} (R_t^{(S)} - \bar{R}^{(S)})^2}{T}}
	\end{equation}
	
	\item Information Ratio: is the average annualized return of a strategy earned in excess of a particular benchmark per unit of risk (measured in terms of standard deviation):
	\begin{equation}
	IR = \frac{\bar{R}^{(S)} - \bar{B}}{\sigma_{R^{(S)}}}
	\end{equation}
	
	where $\bar{B}$ is the average return of the benchmark (e.g., treasury bond, equity index). In our case, it was already set to the 3-month LIBOR rate (Table \ref{MCCSParametersDataset}). It should be mentioned that Information Ratio makes each strategy performance comparable: since we are adjusting average returns by the risk assumed for each strategy, it removes the leverage component that is magnifying/shrinking the returns provided by a certain strategy. 
	
\end{itemize}

\section{Results and Discussions}

%

\begin{figure}[h!]
	\centering
	\includegraphics[width=\linewidth]{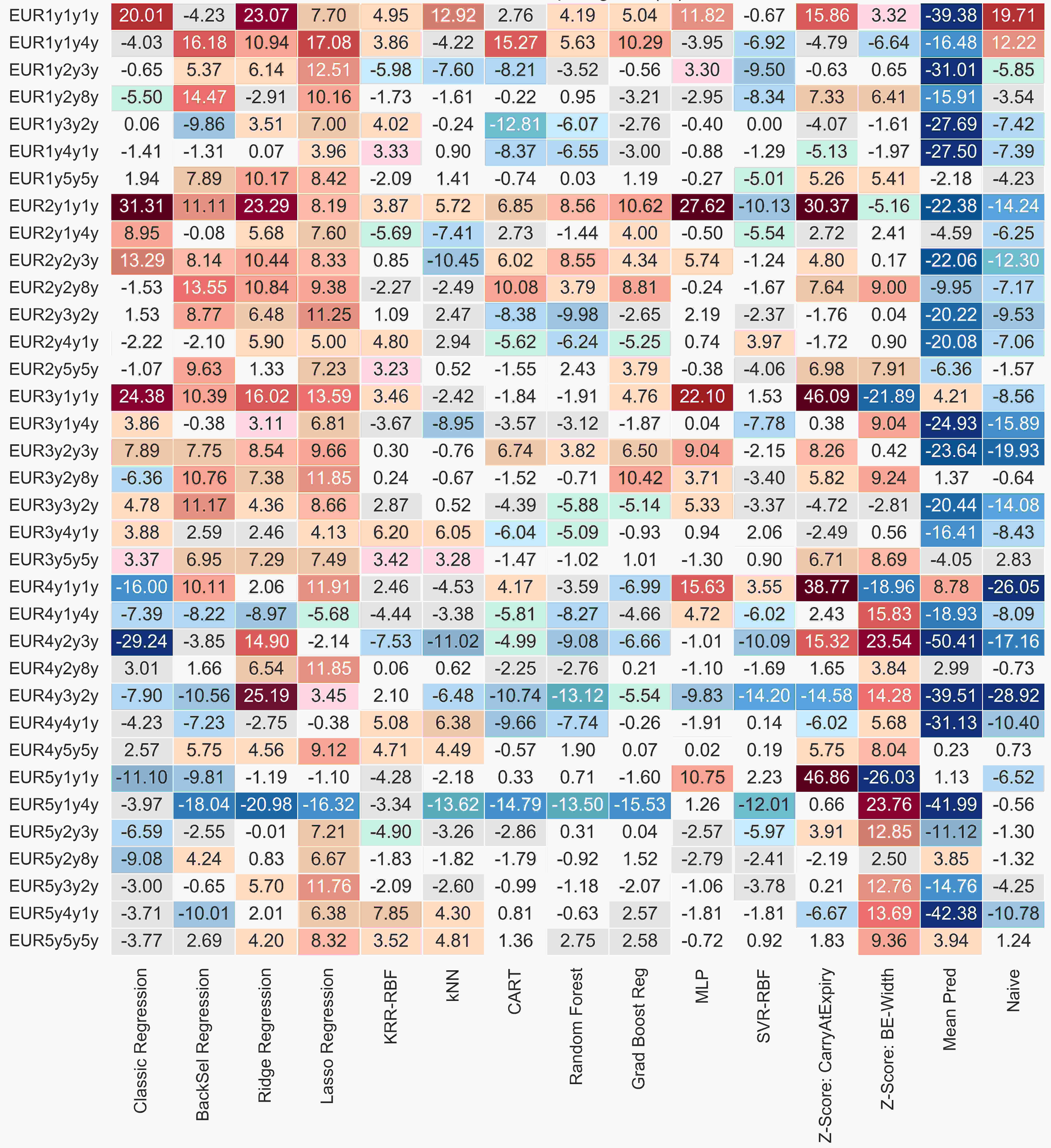}
	\caption{Heatmap with the historical Average Return (\%) during test set.} \label{RSCollectiveHeatMapAvgROI}
\end{figure}

Figure \ref{RSCollectiveHeatMapAvgROI} displays a heatmap with the results of all models for each trade regarding Average Return (\%). Similarly, different remarks can be made over the global picture: (i) the Naive and Mean Pred models underperformed, but the traders benchmarks did perform reasonably well, surpassing the predictive models in many occasions; (ii) from the linear regression family, Lasso Regression followed by Ridge Regression are the ones that performed better; (iii) most nonlinear models failed to provide a decent average return; and (iv) MLP fared well for the trades in the EUR Xy1y1y range, but did not repeat a more stable across other trades.  

\begin{figure}[h!]
	\centering
	\includegraphics[width=\linewidth]{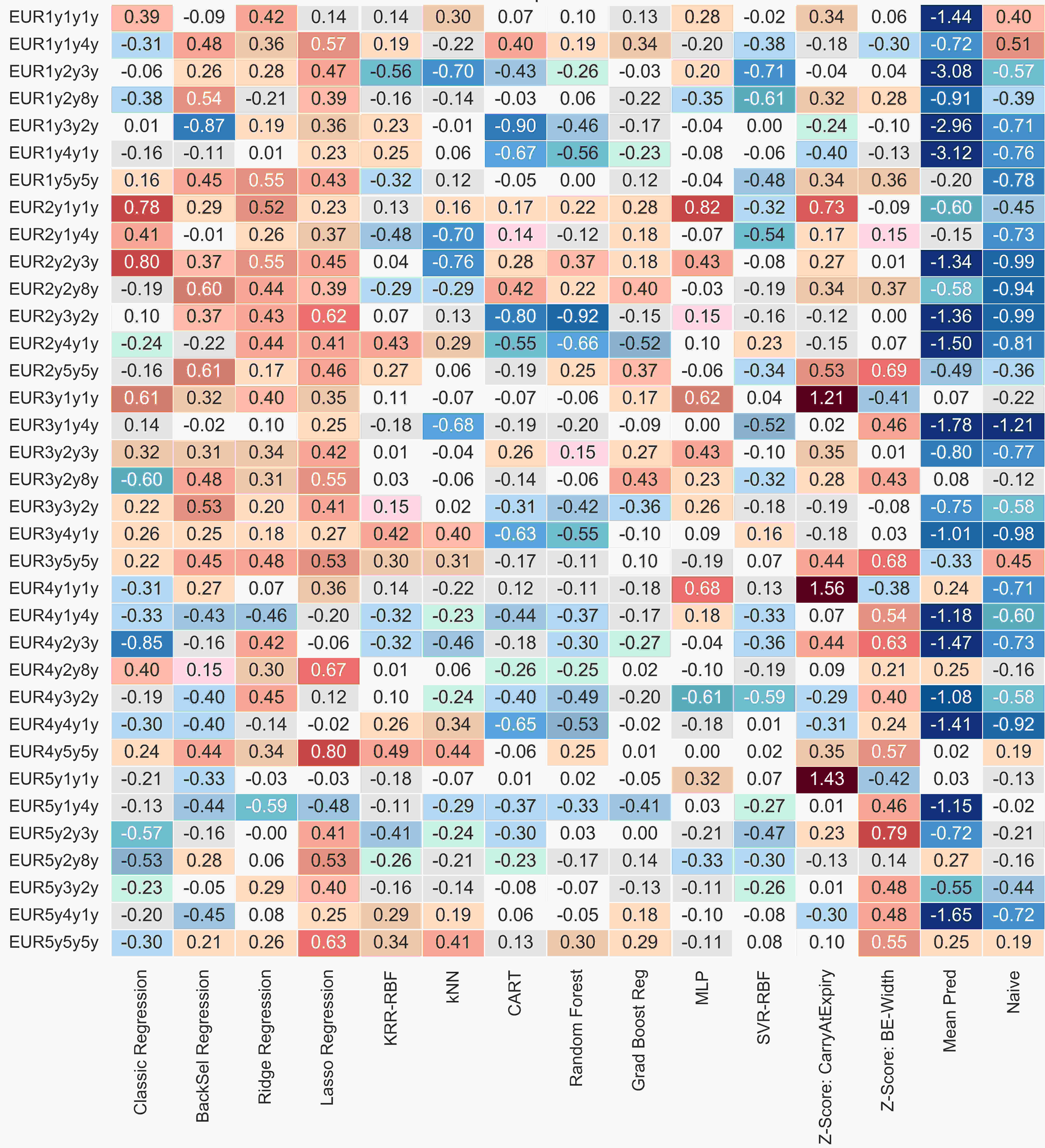}
	\caption{Heatmap with the Information Ratio during the test set.} \label{RSCollectiveHeatMapInformationRatio}
\end{figure}

When we take into account the variability seen in the stream of returns generated by the recommendation system, we may encounter a different picture. Figure \ref{RSCollectiveHeatMapInformationRatio} shows a heatmap with the Information Ratio for all the available combinations of models and trades. 

In general, the models kept their positions unaltered in comparison to the Average Return (\%) -- linear models still fared better than the nonlinear ones --, but now all are standing on a similar scale. Based on these Information Ratio results, Table \ref{RankFriedmanHolmIR} presents a statistical analysis using the average ranks\footnote{When we rank the models for a single MCCS, it means that we sort all them in such way that the best performer is in the first place (receive value equal to 1), the second best is positioned in the second rank (receive value equal to 2), and so on. We can repeat this process for all trades and compute metrics, like the average rank (e.g., 1.35 means that a particular model was placed mostly near to the first place).}, Friedman test and Holm posthoc procedure \cite{Paper:DerracStatComp:2011}.  

\begin{table*}[h!]
	\centering
	\caption{Average ranks, Friedman and Holm post-hoc statistical tests and analysis for Information Ratio.} \label{RankFriedmanHolmIR}
	\scriptsize
	\begin{tabular}{c|cccc}
		\hline
		\hline
		Model & Avg Rank & Z-score & p-value & Holm Correction \\
		\hline
		Mean Pred & 12.86 & 9.63 & \textbf{$<$0.0001} & 0.0036 \\
		Naive & 11.89 & 8.66 & \textbf{$<$0.0001} & 0.0038 \\
		SVR-RBF & 10.60 & 7.37 & \textbf{$<$0.0001} & 0.0042 \\
		CART & 10.11 & 6.89 & \textbf{$<$0.0001} & 0.0045 \\
		Random Forest & 9.31 & 6.09 & \textbf{$<$0.0001} & 0.0050 \\
		kNN & 8.23 & 5.00 & \textbf{$<$0.0001} & 0.0056 \\
		Classic Regression & 8.17 & 4.94 & \textbf{$<$0.0001} & 0.0063 \\
		Grad Boost Reg & 7.69 & 4.46 & \textbf{$<$0.0001} & 0.0071 \\
		KRR-RBF & 7.49 & 4.26 & \textbf{$<$0.0001} & 0.0083 \\
		MLP & 7.20 & 3.97 & \textbf{$<$0.0001} & 0.0100 \\
		BackSel Regression & 6.37 & 3.14 & \textbf{0.0008} & 0.0125 \\
		Z-Score: CarryAtExpiry & 6.09 & 2.86 & \textbf{0.0021} & 0.0167 \\
		Z-Score: BE-Width & 5.91 & 2.69 & \textbf{0.0036} & 0.0250 \\
		Ridge Regression & 4.86 & 1.63 & 0.0517 & 0.0500 \\
		Lasso Regression & \textbf{3.23} & - & - & - \\
		\hline
		Friedman Chi-Square & 117.22 & \textbf{$<$0.0001} &  &  \\
		\hline
		\hline
	\end{tabular}
\end{table*}

\begin{figure}[h!]
	\centering
	\includegraphics[width=\linewidth]{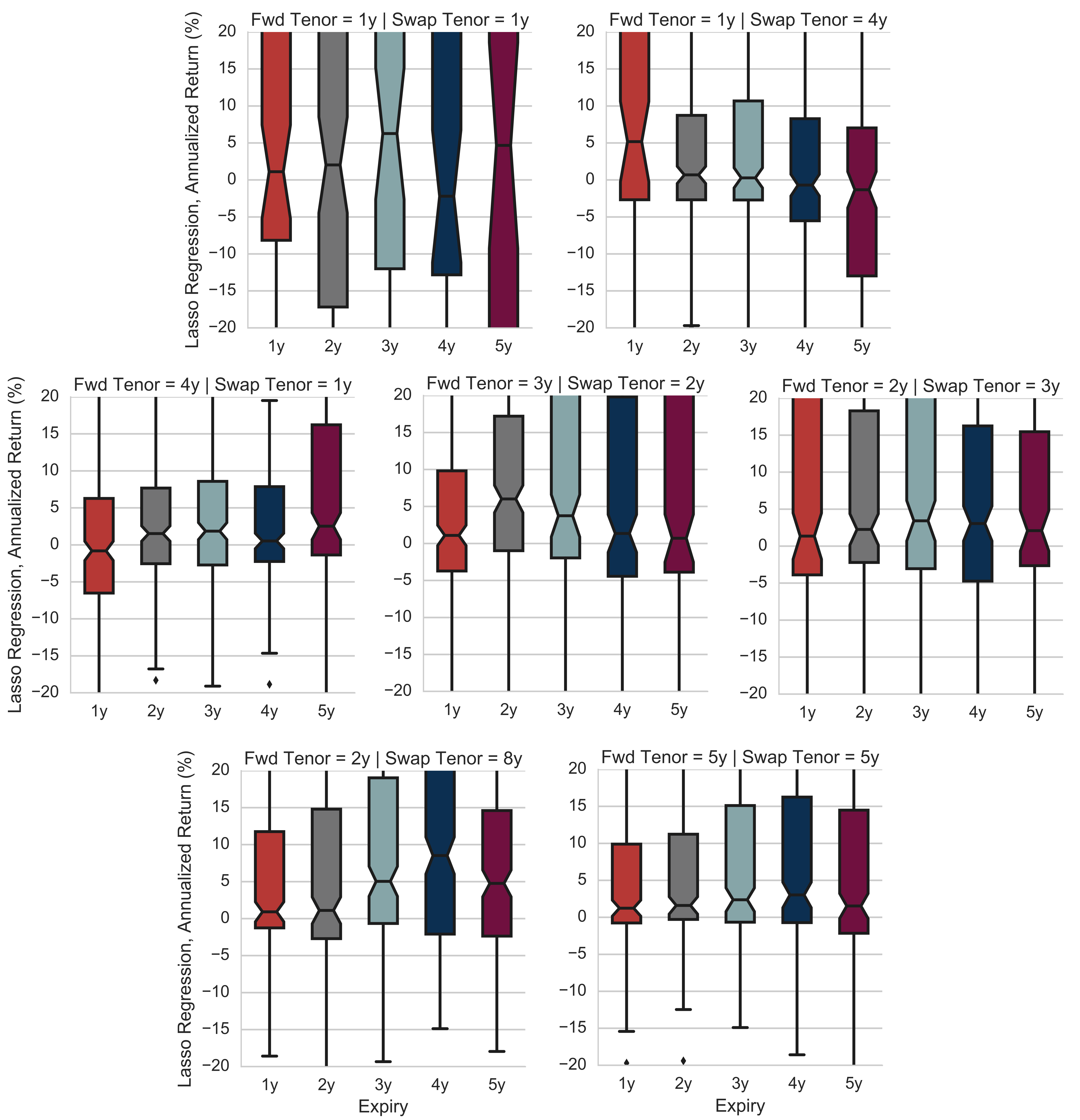}
	\caption{Boxplot with the returns obtained from Lasso Regression for each instrument. All y-axes were fixed between 0.3 and -0.3 (30\% and -30\% annualised return) to facilitate their visualisations and comparisons.} \label{LassoRegressionMCCSBoxReturns}
\end{figure}

When we look at the average rank, Lasso Regression was the top positioned (3.23) while Mean Pred remained most of the time as the worst choice (12.86). The trader's benchmarks performed pretty well, being placed in the third and fourth places. When we compare whether such result fared by Lasso Regression was substantially different from Ridge Regression (4.86), we arrive with a Z-score equal to 1.63 and a p-value of 0.0517. If we set our initial significance level as 0.05 and correct using the Holm procedure (last column) we can assert that Lasso did not perform significantly different from Ridge Regression, but way better than the other models. Therefore, Lasso Regression is capturing some information beyond that is being spanned by the trader's benchmarks, as well as beating almost all other predictive models for this particular task.
Our throughout analysis suggests that Lasso Regression seems to provide in general the best results across a range of metrics and criteria. Figure \ref{LassoRegressionMCCSBoxReturns} uses boxplots as a visualisation tool to decompose the aggregated results shown before, by informing per trade the returns obtained from using the Lasso Regression trading recommendation system.

\begin{figure}[h!]
	\centering
	\includegraphics[width=\linewidth]{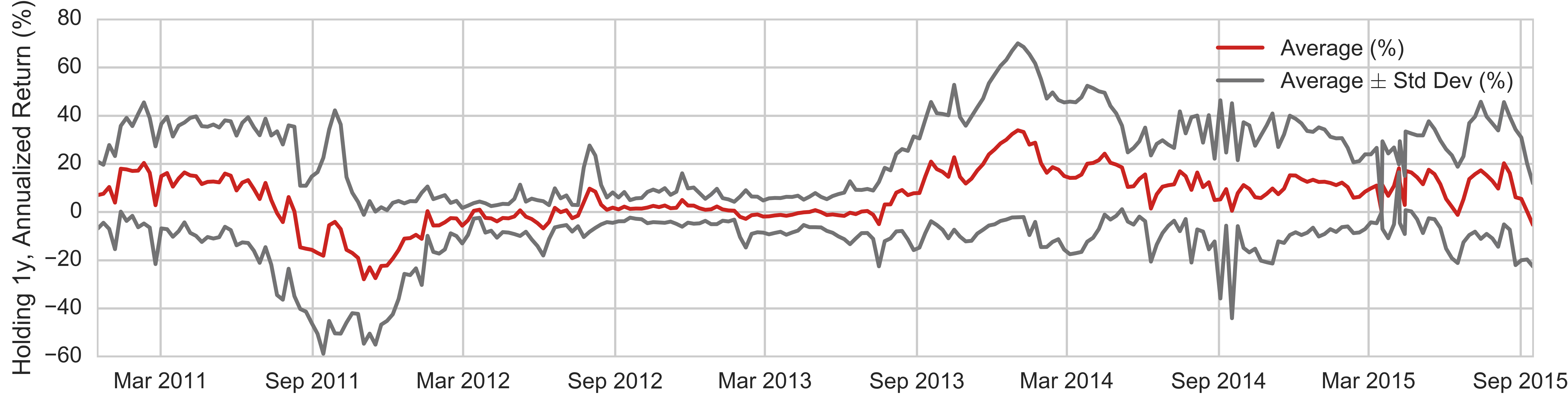}
	\includegraphics[width=\linewidth]{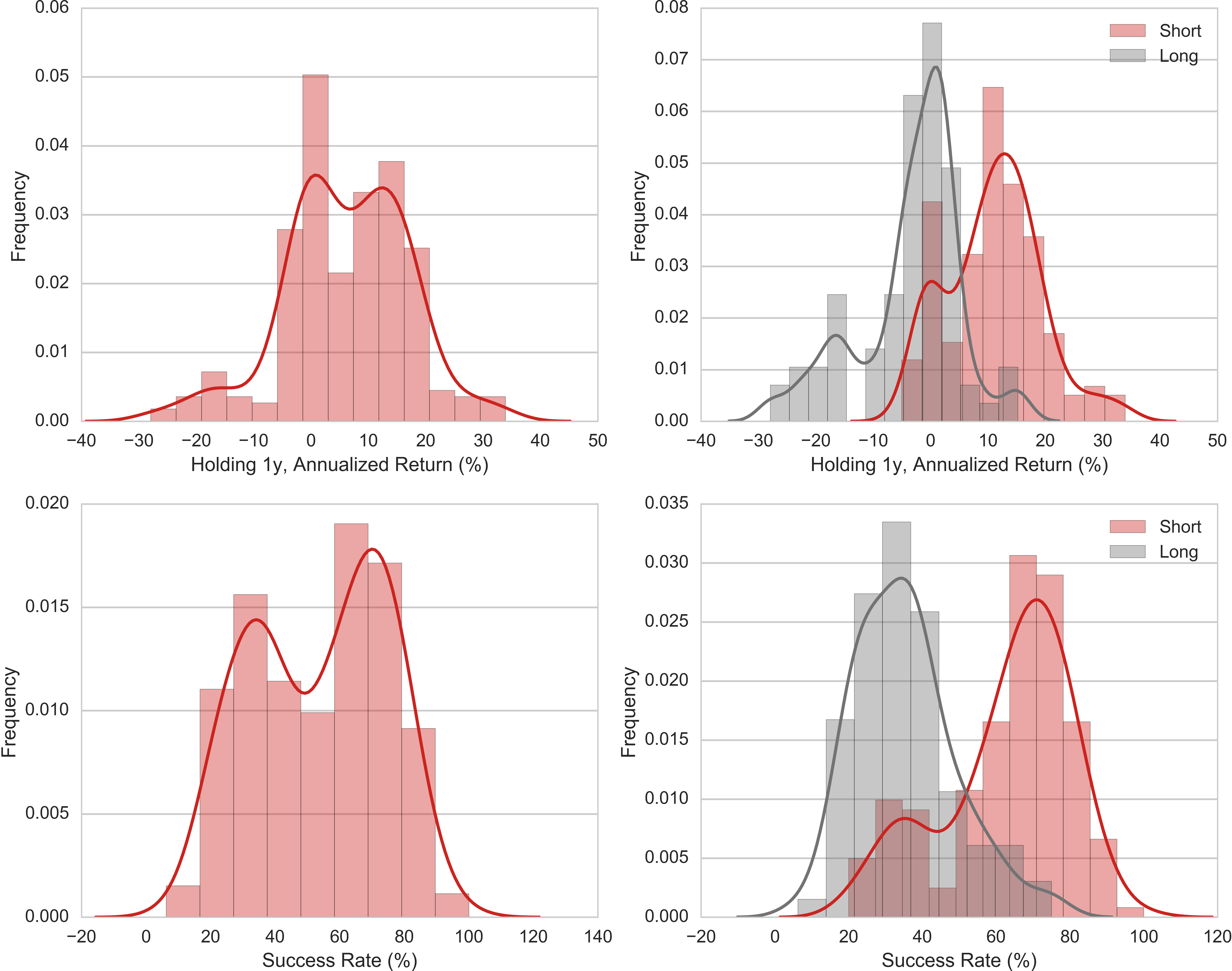}
	\caption{Aggregated returns over the period and histogram of aggregated returns and success rate from trades suggested using Lasso Regression.} \label{LassoRegressionRecSystemHistAgg20062015}
\end{figure}

Overall, some patterns can be spotted from the boxplots: (i) in general the medians are located above zero, meaning that more than half of the trades tended to yield positive returns; (ii) Lasso Regression is exploring long/short position regardless of the most frequent outcome for each MCCS tenure; (iii) it tended to perform well for EUR 3yXyXy trades and since these tended to be historically a challenging pick (medians are centred to zero in these trades), it means that the model is actually capturing some signal from the data and not naively guessing long/short positions; and (iv) the returns distribution, mostly with higher forward and swap tenure (second and third row), tended to be right-skewed -- because the third quartile is far from the median, whereas the first is squeezed towards the median. This last fact denounces that Lasso Regression frequently generates small negative outcomes, and dangerous scenarios are not as likely, which tends to be a desired property for quantitative strategies in general. 
Given that, we now look at the aggregated returns harvested by Lasso Regression during its test phase. These results are consolidated in Figure \ref{LassoRegressionRecSystemHistAgg20062015}, where: (i) the top plot shows the average return with standard deviations obtained across all MCCS per trading week; (ii) the middle reveal histograms, where the left one represents the returns obtained across all MCCS regardless of the trading date, while the right ones displayed the same data but conditioned per position; and (iii) finally the bottom image presents the trading success rate for each long/short position suggested by the model (left), with the break down by long/short position displayed as well (right). To clarify, trading success in this context means being long/short when the returns of trade were positive/negative regardless of its magnitude.
In respect to the top image, we can see that Lasso Regression started well during the first year but suffered a drawdown in the second to third year. This period was marked by higher volatility, mainly due to the final developments of the Euro Crisis period (2010-2012). However, from the third year onwards the average returns always scored positive values, usually ranging from 10\% to 20\% in average. Such performance can be seen stamped on the middle left histogram, where the bulk of returns lies above zero, and not only that but concentrated close to 15\%. This performance was largely generated by Lasso Regression suggesting short positions (middle right), while the long positions were not so successful. Such pattern can be better seen in the histogram located at the bottom, where a bimodal distribution for trading recommendation success rate is depicted. 
Probably the verified outperformance coming from taking short positions in the MCCS is linked with betting against the volatility/variance risk-premium trade \cite{Paper:Choi2017BondVRP:2017}. Roughly this strategy harvest the premium paid by a counterpart for the insurance on large swings in the market (almost the same as selling a put for equities options). Since in general, the market tends to remain range-bounded, the investor shorting the trade can repurchase it later for a smaller premium, profiting from this differential. Lasso Regression did dynamically the opposite and profited from it, largely because in this last 5-6 years was populated of higher volatility periods and tail events.

\begin{figure}[h!]
	\centering
	\includegraphics[width=\linewidth]{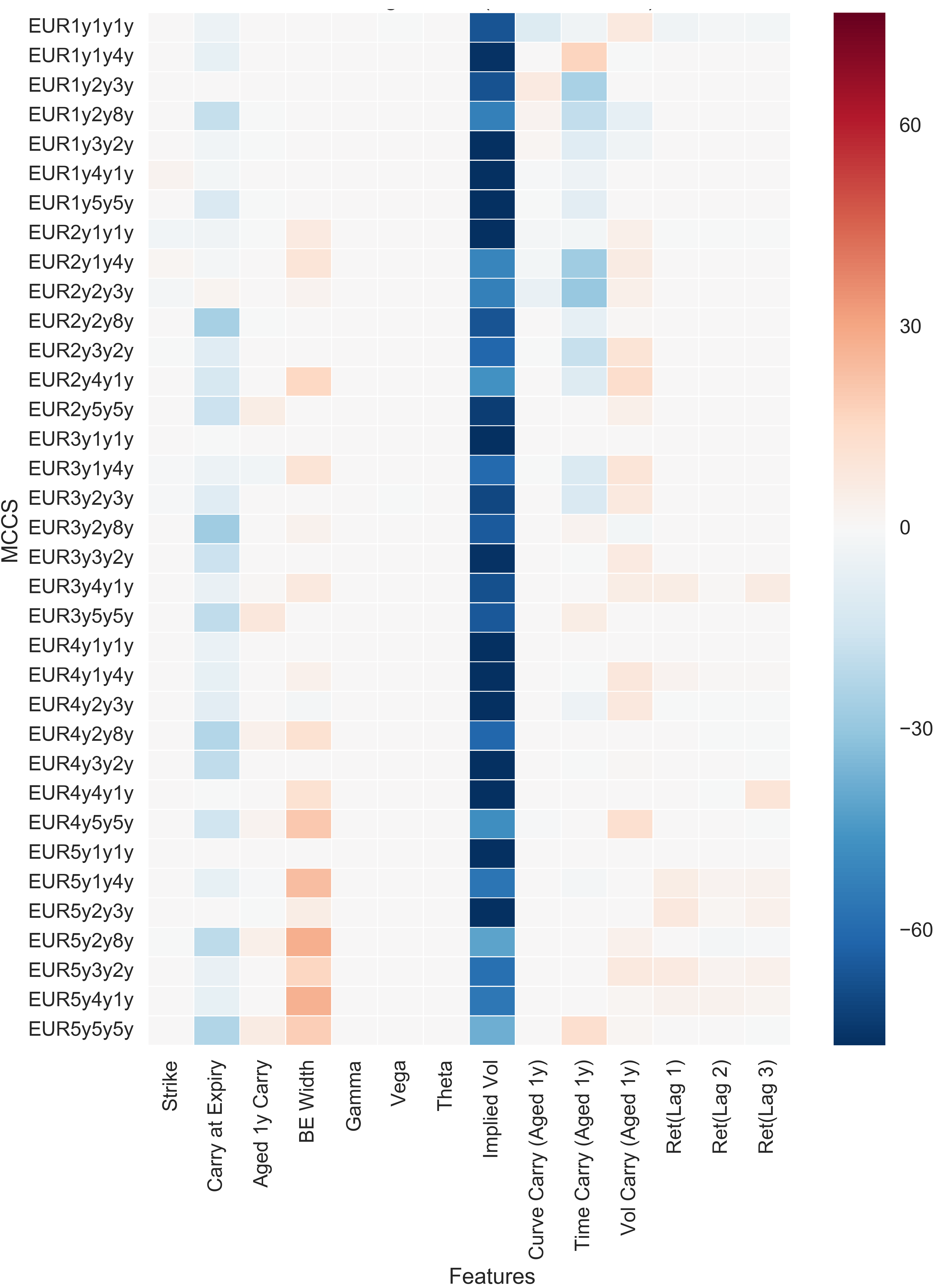}
	\caption{Heatmap with the Feature Significance (\%) obtained from normalised t-stats of Lasso Regression coefficients.} \label{RSEURFeatureRelevanceRandomForest}
\end{figure}

Figure \ref{RSEURFeatureRelevanceRandomForest} help us to analyse which features are being most significant by Lasso Regression for each particular trade. 

\begin{figure}[h!]
	\centering
	\includegraphics[width=0.75\linewidth]{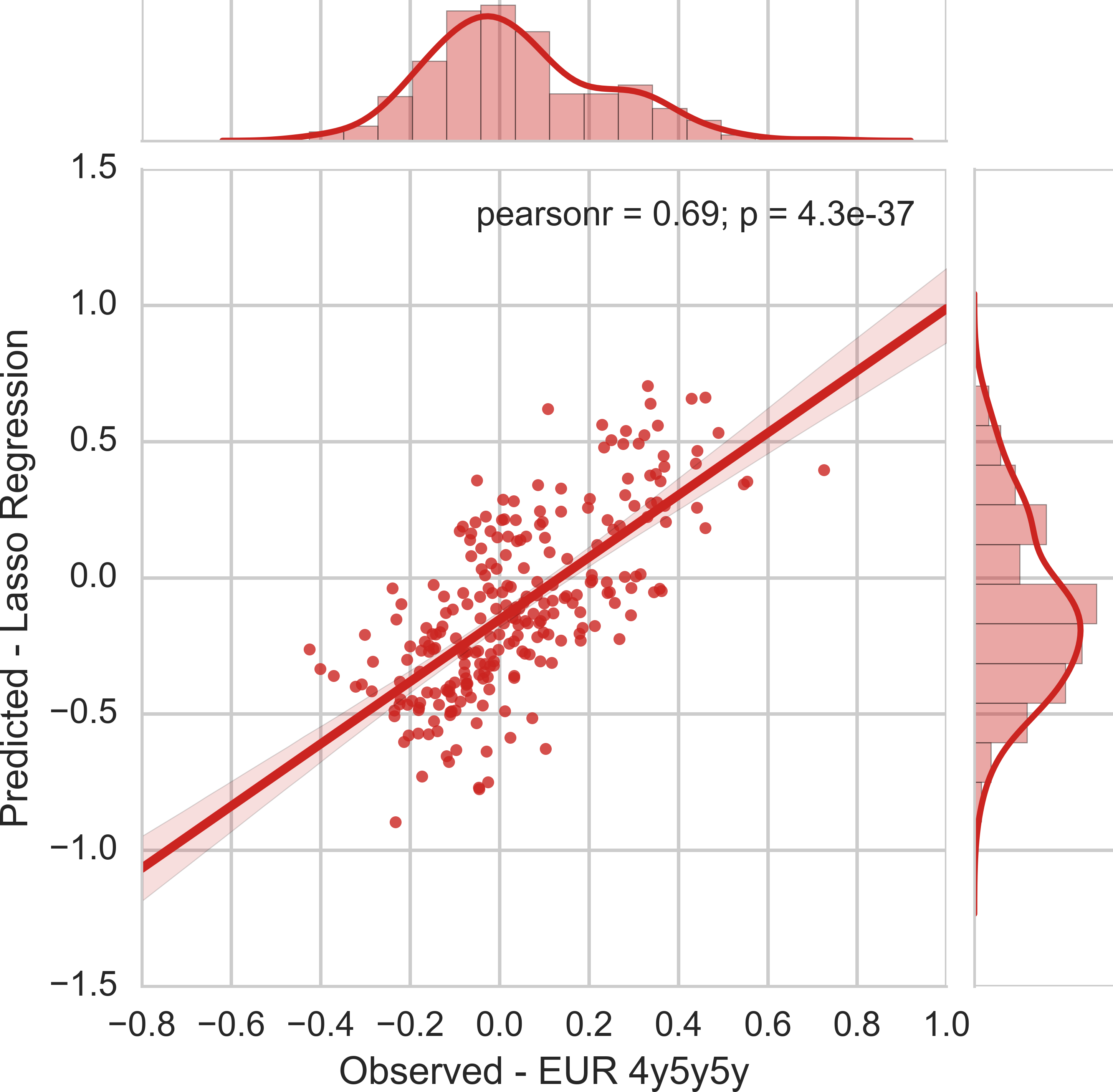}
	\includegraphics[width=0.8\linewidth]{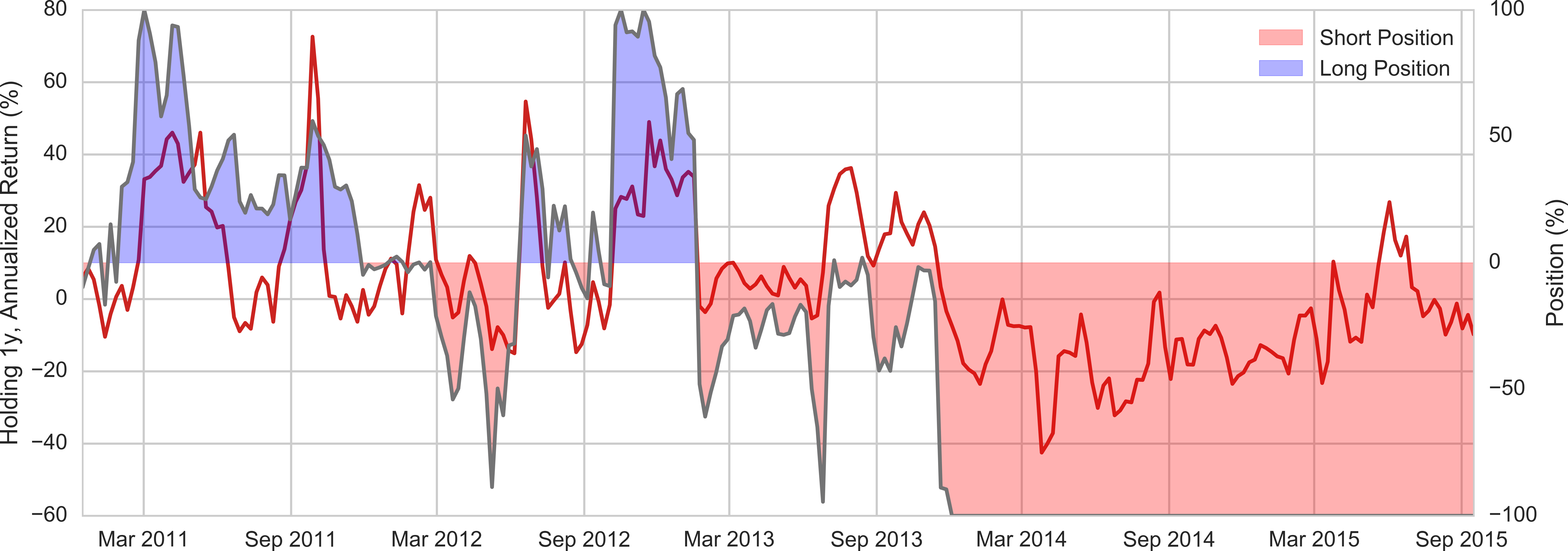}
	\includegraphics[width=0.8\linewidth]{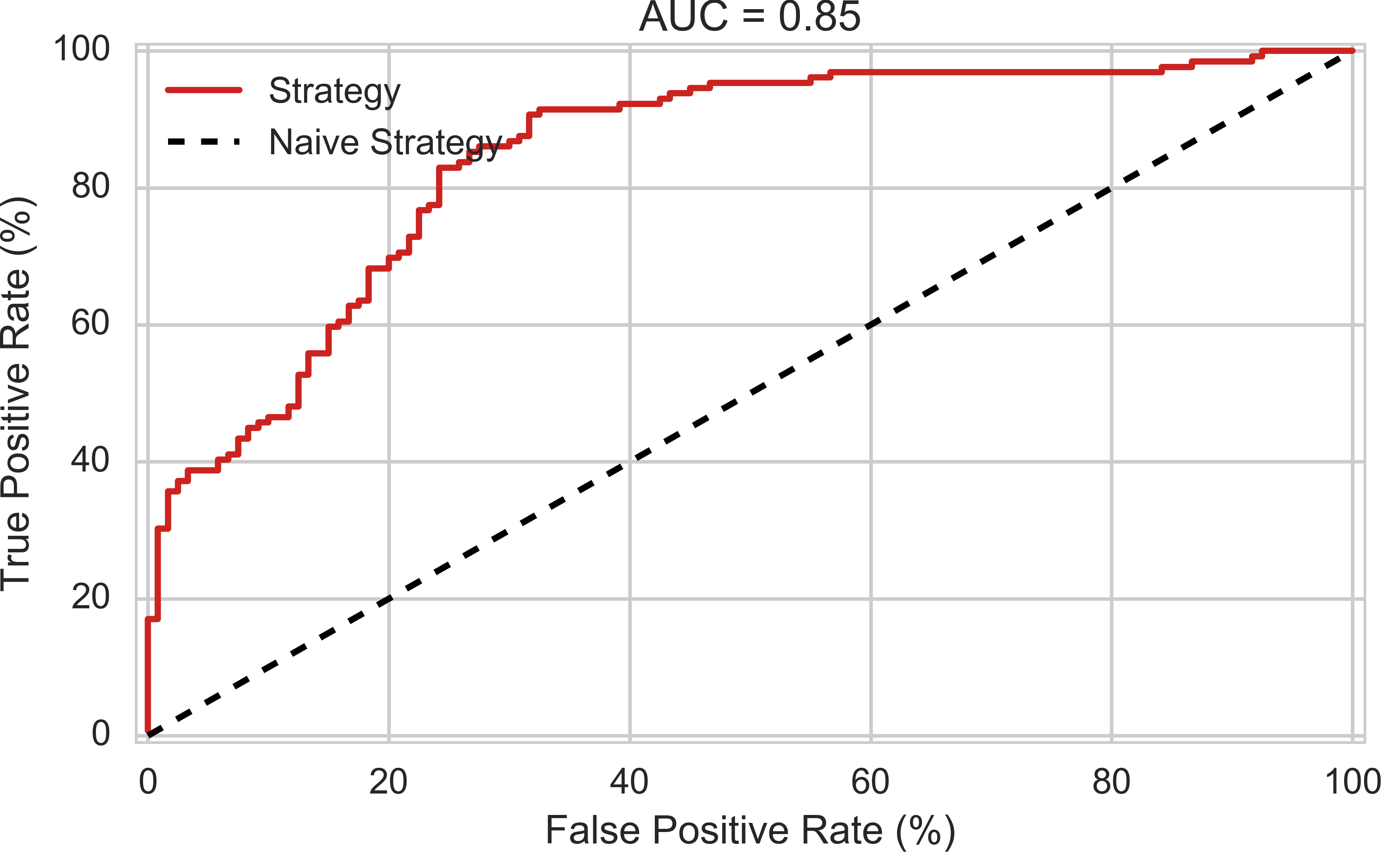}
	\caption{Lasso Regression results for EUR 4y5y5y: predicted versus observed values, returns and long/short signals over the period and receiving operating characteristic curve based on the success rate.} \label{LassoResultsEUR4y5y5y}
\end{figure}

Each cell corresponds to a normalised t-stats\footnote{By normalised t-stats we mean dividing each coefficient t-stat by the sum of the absolute values of all t-stats in the model. The result is a number between -1 and +1, indicating the significant magnitude in comparison to other variables, as well as the direction in which it affects the model predictions. We multiplied it by a one hundred just to work on a more convenient scale of -100\% and 100\%.} from the model coefficients built in the last step from k-rolling-cv. Implied Vol was the most significant feature pointed out by Lasso Regression, is negatively related with the MCCS returns. Other important features were the BE Width -- slightly positively correlated with returns -- and the Carry at Expiry -- negatively related, but probably due to the depressed levels of carry that has been seen in the last batch of data. Lasso Regression promoted in general very sparse models, being the other features playing specific roles for some trades like Time Carry for short-dated trades. Finally, the lagged returns were just relevant for few trades, and perhaps could be omitted for certain trades in the future to guarantee a broader dataset.
We close this section showing results of Lasso Regression for a specific trade: EUR 4y5y5y -- Figure \ref{LassoResultsEUR4y5y5y}. 

Starting from the top, it can be seen that the Lasso Regression was able to predict reasonably well the observed returns after 1 year holding this trade. It is not perceived any over/underestimation of values, with the mirror-shape format indicating a good fit. This observation is reflected in the middle image, where the long/short positions track well the observed returns of the EUR 4y5y5y. The main mistakes are due to events that were not incorporated into the model and perhaps were also unforecastable: (i) March to April 2012, possibly due to the Greek Debt Restructuring Agreement; (ii) May to November 2013, linked with the Taper Tantrum event in the US and its effect on Europe rates. 

Although such events have influenced in the strategies return, last plot shows that the trading success of Lasso Regression has attained a historical Area Under the Curve (AUC) of 0.85, with the capacity to control the false alarms to 5\% and still recommend trades accurately 40\% of the time. This is a good indicator since in the onset of the recommendation system it is better to reduce the chances of suggesting a bad trade than missing a good pick.

\pagebreak
\clearpage
\section{Conclusions}

This work proposed a trading recommendation system for Mid-Curve Calendar Spread Trades (MCCS). We proposed a recommendation system that could analyse and rank a set of fixed income derivatives trades. Our first experiment is designing and applying this method for Mid-Curve Calendar Spread trades. Therefore, we started the methodology by showing the dataset: it comprised of 35 MCCS trades, ranging from September 2006 to September 2016, with different expirations, forward and swap tenures. For each particular trade, we described how the sampling of inputs (metrics, sensitivities and lagged returns) and outputs (returns from unwinding the trade after one year of its start) were computed on a weekly basis. Then, we displayed the modelling strategy by highlighting the models that were trained as well as which hyperparameters were investigated during the nested resampling step. Before entering the results section, we presented the backtesting setting with the performance measures used to compare different methodologies.

Most models provided results better than the modelling benchmarks (Mean and Naive), yet very few were able to outperform the trader's benchmarks. Our results suggested that linear models with shrinkage procedures (e.g., Ridge and Lasso) tended to perform better than their nonlinear counterparts (like Kernel Ridge Regression, SVR and MLP). Also, regarding interpretability, they tend to be easier to convey to the traders, since most are versed in linear models. When we delved into Lasso Regression results, we found out that this model wielded some interesting features like: (i) it learned a type of volatility buying/selling strategy without being programmed to do so; (ii) its returns distribution across all MCCS tended to be right-skewed, meaning that we are more hedged towards dangerous scenarios with greater chances of upsides; (iii) it matched traders view on selecting good trades, but adding some dynamic view on it since Carry at Expiry is now negatively linked with returns, rather than the original view from the traders. We believe that Lasso Regression will be our choice for a first version of the trading recommendation system, with future developments giving space to different models and mixed approaches.

\section*{Acknowledgment}
Adriano Soares Koshiyama wants to acknowledge the funding for its PhD studies provided by the Brazilian Research Council (CNPq) through the Science Without Borders program. Also, the authors would like to thanks, Guillaume Andrieux, Tomoya Horiuchi, Gerald Rushton, Tam Rajendran, and Anthony Morris for all the comments and support during this research.

\balance
\bibliographystyle{plain}
\bibliography{bibthesis}

\end{document}